\documentclass[aps,prl,twocolumn,showpacs,groupedaddress,nofootinbib]{revtex4}  
\usepackage{graphicx}  
\usepackage{dcolumn}   
\usepackage{bm}        
\usepackage{amssymb}   

\def\MET{{\mbox{$E\kern-0.57em\raise0.19ex\hbox{/}_{T}~$}}}
\def\METnoSpace{{\mbox{$E\kern-0.57em\raise0.19ex\hbox{/}_{T}$}}}
\newcommand{\simleq}{\; \raisebox{-0.4ex}{\tiny$\stackrel{{\textstyle<}}{\sim}$}\;}
\def\lsim{\mathrel{\rlap{\lower4pt\hbox{\hskip1pt$\sim$}}\raise1pt\hbox{$<$}}}
\def\gsim{\mathrel{\rlap{\lower4pt\hbox{\hskip1pt$\sim$}}\raise1pt\hbox{$>$}}}
\begin{document}


\hspace{5.2in} \mbox{Fermilab-Pub-07/560-E}

\title{Search for Supersymmetry in Di-Photon Final States at $\sqrt{s}$ = 1.96 TeV}
%
\author{V.M.~Abazov$^{36}$}
\author{B.~Abbott$^{76}$}
\author{M.~Abolins$^{66}$}
\author{B.S.~Acharya$^{29}$}
\author{M.~Adams$^{52}$}
\author{T.~Adams$^{50}$}
\author{E.~Aguilo$^{6}$}
\author{S.H.~Ahn$^{31}$}
\author{M.~Ahsan$^{60}$}
\author{G.D.~Alexeev$^{36}$}
% ...

\affiliation{\vspace{0.1 in}(The D\O\ Collaboration)\vspace{0.1 in}}
% ...
\affiliation{$^{6}$University of Alberta, Edmonton, Alberta, Canada,
                Simon Fraser University, Burnaby, British Columbia, Canada,
                York University, Toronto, Ontario, Canada, and
                McGill University, Montreal, Quebec, Canada}
% ...
\affiliation{$^{29}$Tata Institute of Fundamental Research, Mumbai, India}
% ...
\affiliation{$^{31}$Korea Detector Laboratory, Korea University, Seoul, Korea}
% ...
\affiliation{$^{36}$Joint Institute for Nuclear Research, Dubna, Russia}
% ...
\affiliation{$^{50}$Florida State University, Tallahassee, Florida 32306, USA}
% ...
\affiliation{$^{52}$University of Illinois at Chicago,
                Chicago, Illinois 60607, USA}
% ...
\affiliation{$^{60}$Kansas State University, Manhattan, Kansas 66506, USA}
% ...
\affiliation{$^{66}$Michigan State University,
                East Lansing, Michigan 48824, USA}
% ...
\affiliation{$^{76}$University of Oklahoma, Norman, Oklahoma 73019, USA}
% ...
\date{October 21,2007}

\begin{abstract}
We report results of a search for supersymmetry (SUSY) with gauge-mediated symmetry
breaking in di-photon events collected by the D0 experiment at the Fermilab Tevatron 
Collider in 2002--2006. In 1.1 fb$^{-1}$ of data, we find no significant excess beyond the background
expected from the standard model and set the most stringent lower limits to date
for a standard benchmark model on the lightest neutralino and chargino masses
of 125 GeV and 229 GeV, respectively, at 95\% confidence.
\end{abstract}

\maketitle
Low-scale SUSY is one of the most promising solutions to the hierarchy
problem associated with the intrinsic disparity between the electroweak and Planck
scales. It postulates that for each
known particle there exists a superpartner, thereby stabilizing the radiative corrections to the Higgs boson mass. 
Bosons have fermion superpartners, and vice versa. None of the superpartners have yet been observed,
and superpartner masses must therefore be much larger than those of their partners,
$i.e.$, SUSY is a broken symmetry. Experimental signatures of supersymmetry
are determined through the manner and scale of SUSY breaking. In models with
gauge-mediated supersymmetry breaking (GMSB)~\cite{gmsb_original,gmsbsusy},
it is achieved through the introduction of new chiral supermultiplets, called
messengers that couple to the ultimate source of supersymmetry breaking
and to the SUSY particles. At colliders, assuming $R$-parity
conservation~\cite{rpar}, superpartners are produced in pairs
($\tilde{\chi}_1^+\tilde{\chi}_1^-$ and $\tilde{\chi}_1^\pm\tilde{\chi}_2^0$ production dominates in most cases)
and decay to the standard model particles and next-to-lightest SUSY particle (NLSP), which can be
either a neutralino or a slepton. In the former case, which is considered
in this note, the NLSP decays into a photon and a gravitino (the lightest
superpartner in GMSB SUSY models, with mass less than $\approx$ 1 keV).
The gravitino is stable, and escapes detection, creating an apparent imbalance 
in transverse momentum (\METnoSpace) in the event. 
GMSB SUSY final states are therefore characterized by 
two energetic photons and large missing transverse momentum.
The differences in event kinematics between particular GMSB SUSY models
result in slightly different experimental sensitivities~\cite{gmsb_original},
and to obtain a quantitative measure of limits on SUSY we consider a model 
referred to as ``Snowmass Slope
SPS 8''~\cite{snowmass}. This model has only a single dimensioned parameter:
an energy scale $\Lambda$ that determines the effective scale of SUSY breaking. The
minimal GMSB parameters correspond to a messenger mass $M_m = 2\Lambda$,
the number of messengers $N_5 = 1$, the ratio of the vacuum expectation
values of the two Higgs fields tan$\beta$ = 15, and the sign of the Higgsino
mass term $\mu > 0$. The neutralino lifetime is not defined within the model.
For this analysis, it is assumed to be sufficiently short to yield decays
with prompt photons.

Searches for GMSB SUSY were carried out by collaborations at the CERN LEP
collider~\cite{lep} and at the Fermilab Tevatron collider in both
Run I~\cite{tevrun1} and early in Run~II~\cite{d0-tevrun2, cdf-tevrun2}. 
The initial limits from CDF and D0 for Run~II, based on the SPS 8 model, were 
combined~\cite{tevcombined} to yield $\Lambda > 84.6$~TeV 
corresponding to the limit on the chargino mass of 209~GeV, at 95\% confidence.
Complementary searches for GMSB SUSY with R-parity violation were performed by the H1 
experiment at HERA \cite{h1-gmsb}.

This analysis is an update of that described in Ref.~\cite{d0-tevrun2},
with about a factor of three more data and improved photon identification based on: (i)
an electromagnetic (EM) cluster "pointing" algorithm that
predicts the origin of a photon with a resolution
of about 2~cm along the beam axis, thereby eliminating the largest instrumental
background associated with misreconstruction of the primary
interaction vertex, and (ii) an improved track veto requirement
that suppresses sources of background with electrons
in the final state. We also use an improved likelihood
fitter~\cite{likelihoodFitter} to set limits on the
scale parameter~$\Lambda$.

The data in this analysis were recorded using single EM triggers with the D0 detector~\cite{d0det},
the main components of which are an inner tracker,
liquid-argon/uranium calorimeters, and a muon spectrometer.
The inner tracker consists of silicon microstrip and central scintillating-fiber trackers 
located in a 2~T superconducting solenoidal magnet, providing
measurements up to pseudorapidities
\footnote{Pseudorapidity is defined as 
$-\log(\tan(\frac{\theta}{2}))$, where $\theta$ is the angle between the 
particle and the proton beam direction.} 
of $|\eta|\approx3.0$ and $|\eta|\approx1.8$,
respectively.
The calorimeters are finely segmented
and consist of a central section (CC) covering $|\eta|<1.2$ and
two endcap calorimeters extending coverage to $|\eta|\approx 4$,
all housed in separate cryostats~\cite{d0cal}. The electromagnetic
section of the calorimeter has four longitudinal layers and
transverse segmentation of 0.1 $\times$ 0.1 in $\eta - \phi$ space
(where $\phi$ is the azimuthal angle), except in the third layer,
where it is 0.05 $\times$ 0.05.
The central preshower (CPS) system is placed between the solenoid and the
calorimeter cryostat and covers $|\eta| \simleq 1.2$. The CPS provides precise 
measurement of positions of EM showers. The axes of EM showers are reconstructed
by fitting straight lines to shower positions measured in the four longitudinal 
calorimeter layers and the CPS (EM "pointing").
The data for this study were collected between 2002 and summer 2006, 
using inclusive single EM triggers that are almost 100\% efficient to 
select signal data. The integrated luminosity~\cite{d0lumi} of the 
sample is $1100\pm70$~pb$^{-1}$.

Photons and electrons are identified based on reconstructed EM
clusters using calorimetric information and further classified
into electron and photon candidates, based on tracking information.
The EM clusters are selected from calorimeter
clusters using the simple cone method (of radius
${\cal R} = \sqrt{(\Delta\eta)^{2} + (\Delta\phi)^{2}} = 0.4$) by
requiring that (i) at least 90\% of the energy is deposited in the
EM section of the calorimeter, (ii) the calorimeter isolation variable
$I = [E_{\rm tot}(0.4) - E_{EM}(0.2)]/E_{EM}(0.2)$ is less than 0.07, 
where $E_{\rm tot}(0.4)$ is the total shower energy in a cone of radius ${\cal R}=0.4$, and
$E_{EM}(0.2)$ is the EM energy in a cone of radius ${\cal R}=0.2$, (iii) the
transverse, energy-weighted, width of the EM cluster in the third
EM calorimeter layer is smaller than 0.04 rad, and (iv) the scalar
sum of the transverse momenta ($p_T$) of all tracks originating
from the primary vertex in an annulus of $0.05 < {\cal R} < 0.4$
around the cluster is less than 2 GeV.  The EM cluster is further
defined as an electron candidate if it is spatially matched to
activity in the tracker, and as a photon candidate otherwise.
The tracker activity can be either a reconstructed track or
a density of hits in the silicon microstrip and
central fiber trackers consistent with a track, i.e., an electron.
The latter requirement allows for increasing electron track-matching efficiency, $\epsilon_{\rm trk}$,
measured in $Z\rightarrow ee$ data, from $(93.0\pm0.1)$\% to
$(98.6\pm0.1)$\% by identifying electrons with lost tracks due to
hard bremsstrahlung and/or inefficiency of the inner trackers.
This reduces electron backgrounds to photons 
by a factor of five, while keeping the efficiency of anti-track
activity requirement high. We measure that $(91\pm3)$\% of photon candidates
in $Z\rightarrow ee\gamma$ data satisfy the anti-track activity requirement.

Jets are reconstructed using the iterative, midpoint
cone algorithm~\cite{d0jets} with a cone size of ${\cal R} = 0.5$.
The missing transverse energy is determined from the energy deposited in the calorimeter
for $|\eta| < 4$ and is corrected for the EM and jet energy scales.

We select $\gamma\gamma$ candidates by requiring events to have
two photon candidates, each with transverse energy $E_T > 25$ GeV
identified in the CC with $|\eta| < 1.1$. We require
that at least one of the photon candidates be matched to a CPS
cluster, and that the primary vertex be consistent with that of the
photon candidate (obtained from the EM pointing).
The accuracy of the determination of the photon vertex is measured using
photons from final state radiation in $Z\rightarrow ee\gamma$ data
sample and found to be $2.3 \pm 0.3$ cm. The requirement of
consistency between the photon and primary vertices ensures
correct calculation of the transverse energies and tracking
isolation requirements. The accuracy of primary vertex association
is studied in GMSB SUSY Monte Carlo simulated events, where the primary
vertex is identified correctly in $(98.5\pm0.1)$\% of the events while the
photon vertex matches the primary vertex in $(95.8\pm0.1)$\%.

To reduce potential bias in the measurement of \MET from mismeasurement of jet
transverse momentum, we also require that the jet with the highest $E_T$
(if jets are present in the event) be separated from the \MET
in azimuth by no more than 2.5 radians. This selection yields 2341 events
(the $\gamma\gamma$ sample).

All instrumental backgrounds arise from standard model processes,
with either genuine \MET ($W\gamma$, $W$+jet, and $t\bar{t}$ production)
or without inherent  \MET (direct photon, multi-jet, and $Z\rightarrow ee$
production). All these backgrounds are measured using data.

The former source always has an electron in the final state which
is misidentified as a photon. The contribution of this background
to the \MET distribution in data can be estimated using an
$e\gamma$ sample (selected by requiring an electron and a photon
candidate and using the same kinematical requirements as
for the $\gamma\gamma$ sample) scaled by the probability of an electron-photon
misidentification which is measured using $Z \rightarrow ee$ data.
First, the \MET distribution in the $e\gamma$ sample must be corrected
for the contribution from events with no real \METnoSpace.
The contribution from Drell-Yan events is taken into account by obtaining
the \MET distribution for the $ee$ sample (selected by requiring two electron
candidates and applying the same kinematical requirements as for
the $\gamma\gamma$ sample) which is dominated by Drell-Yan events.
The Drell-Yan \MET distribution is further normalized to the number of
$Z$ boson events in the $e\gamma$ sample (the latter is determined
by fitting the $e\gamma$ invariant mass spectrum to the $Z$ boson mass peak).

The contribution from the multi-jet processes is estimated from a data sample
(referred to as the {\it QCD} sample) selected by requiring two EM clusters that
(a) satisfy all the kinematic selection used to select $\gamma\gamma$
sample and (b) satisfy all the photon identification criteria but fail
the shower shape requirement. The \MET distribution in the {\it QCD} sample
is normalized to the number of the events in the $e\gamma$ sample with
\MET $<$ 12 GeV after subtraction of the Drell-Yan contribution as determined above.
The expected number of $W\gamma$, $W+{\rm jet}$, and $t\bar{t}$ events with
$\MET < 12$ GeV is negligible.

After the Drell-Yan and multi-jet contributions to the $e\gamma$ sample are
subtracted, the resulting \MET distribution is scaled by
$(1-\epsilon_{\rm trk})/\epsilon_{\rm trk}$, where $\epsilon_{\rm trk}$ is the
efficiency of the track-matching requirement to obtain the estimate of \MET
distribution for the background with genuine \MET.

The background from events with no inherent \MET is divided into multi-jet
events with two real isolated photons and events where one or both
photons are misidentified jets. Since the \MET resolution for both sources
is dominated by the photon energy resolution, the \MET distributions for
the two sources are very similar. However, misidentified jets have a
different energy response compared with that of real photons which leads to
a slight difference in the shapes of the  \MET distributions. For the real
di-photon events, the \MET is assumed to have the same shape as that of the Drell-Yan
events. For misidentified jets, the shape of the \MET distribution is taken
from the {\it QCD} sample. Relative normalization of the two sources is
obtained using a fit to the \MET distribution in the $\gamma\gamma$
sample. We check that the fit is not sensitive to possible
signal contribution, and cross-check with a method that
estimates the $\gamma\gamma$ sample purity using the measured shower shape in
the CPS. The relative fraction of di-photons is $(60\pm20)$\% and
this uncertainty is propagated as a systematic
uncertainty for the limit setting. Absolute normalization of the \MET distributions from both sources
is determined so that the number of events with \MET $< 12$ GeV matches that in 
the $\gamma\gamma$ sample.

The largest physics backgrounds are from
$Z\gamma\gamma\rightarrow \nu\nu\gamma\gamma$ and
$W\gamma\gamma\rightarrow \ell\gamma\gamma\nu$ processes. Contributions
from these backgrounds are estimated as $0.15\pm0.06$ and
$0.10\pm0.04$ events, respectively, using  {\sc CompHep}~\cite{comphep}
Monte Carlo simulation, cross-checked with {\sc MadGraph}~\cite{madgraph}.
The contribution of these backgrounds to the  \MET distribution
is taken from Monte Carlo simulation, with number of events normalized
to the integrated luminosity of the data sample.

\begin{figure}
  \includegraphics[width=9cm]{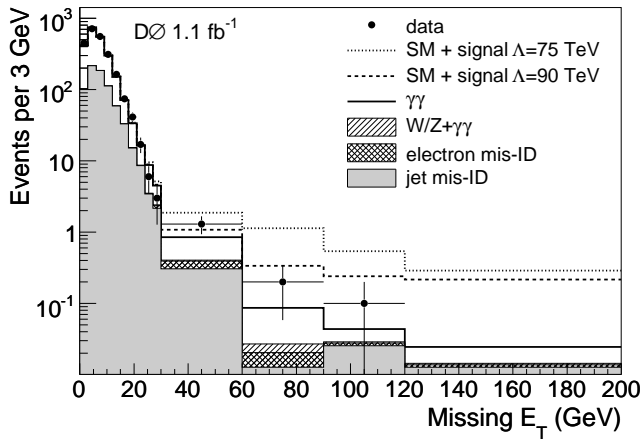}\\
  \caption{The \MET distribution in $\gamma\gamma$ data with $W/Z+\gamma\gamma$ background
  (hatched histogram), instrumental background with no genuine \METnoSpace: $\gamma\gamma$
  (solid black line) and multi-jet (filled histogram), and
  background from processes with genuine \MET and a misidentified electron (cross-hatched
  histogram). The expected \MET distributions if GMSB SUSY events were present are shown
   as dotted and dashed lines.}
\label{fig:ggmet_final_gg}
\end{figure}

The \MET distribution for the $\gamma\gamma$ sample, with contributions 
from physics background ($W/Z + \gamma\gamma$), and instrumental background 
with genuine \MET (processes with misidentified electrons)
and no inherent \MET ($\gamma\gamma$ and multi-jet) is given in
Fig~\ref{fig:ggmet_final_gg}. We also illustrate the  \MET distribution expected from
GMSB SUSY for two values of $\Lambda$. The number of observed events,
as well as expected background and signal from GMSB SUSY
for \MET $> 30$~GeV and $>60$~GeV are given in Table~\ref{tab:gg_summary}.
%
\begin{table*}
\begin{center}
\begin{ruledtabular}
\begin{tabular}{ccccccccc}
 & \multicolumn{4}{c}{Background events} & \multicolumn{2}{c}{Expected signal events} & Observed events \\
                & Genuine \MET  & No \MET       & Physics        & Total
                                                                                & $\Lambda$ = 75 TeV
                                                                                               & $\Lambda$ = 90 TeV
                                                                                                                & \\
                                                                                                                \hline
\MET $>$ 30 GeV & 0.97$\pm$0.12 & 9.62$\pm$1.12 & 0.19$\pm$0.07 & 10.8$\pm$1.1 & 28.3$\pm$4.2 & 8.7$\pm$1.3 & 16 \\
\MET $>$ 60 GeV & 0.11$\pm$0.04 & 1.44$\pm$0.43 & 0.08$\pm$0.04 &  1.6$\pm$0.4 & 18.1$\pm$2.7 & 6.4$\pm$1.0 & 3 \\
\end{tabular}
\caption{\label{tab:gg_summary} Numbers of background events from $W\gamma$, $W+{\rm jet}$, and 
$t\bar{t}$ (Genuine \METnoSpace), no inherent \MET (No \METnoSpace), $Z\gamma\gamma\rightarrow \nu\nu\gamma\gamma$ and
$W\gamma\gamma\rightarrow \ell\gamma\gamma\nu$ (Physics) processes; the total number of expected background events;
numbers of expected GMSB SUSY signal events for two values of $\Lambda$;
and the observed numbers of events for \MET $>$ 30 GeV and 60 GeV. Errors are statistical and systematic combined.}
\end{ruledtabular}
\end{center}
\end{table*}

The expected GMSB signal efficiency is estimated from Monte Carlo simulation generated
for several points on the Snowmass Slope (see Table~\ref{tab:susy}), covering
the neutralino mass range from 170 GeV to 280 GeV. 
Although $\tilde{\chi}_1^+\tilde{\chi}_1^-$ and $\tilde{\chi}_1^\pm\tilde{\chi}_2^0$
processes dominate, we consider all GMSB SUSY production channels. 
We used {\sc ISAJET 7.58}~\cite{isajet} to determine SUSY
interaction eigenstate masses
and couplings. {\sc PYTHIA 6.319}~\cite{pythia} is used to generate the events
after determining the sparticle masses, branching fractions and leading order
(LO) production cross sections using CTEQ6L1 parton distributions~\cite{cteq}.
The generated events are processed through a full
 {\sc GEANT}-based \cite{geant}  
 detector simulation and
the same reconstruction code as used for data. The LO signal cross sections
are scaled to match the next-to-leading order (NLO) prediction using
$k$-factor values (see Table~\ref{tab:susy}), extracted from Ref.~\cite{kfactor}.

The systematic error on the expected number of signal events comes from the uncertainties in photon identification efficiency (10\%), statistics in MC samples (5\%),
track veto requirement (3\%), and  trigger efficiency (4\%). These were obtained using $Z\rightarrow e^+e^-$ and $Z\rightarrow e^+e^-\gamma$ decays in data and in MC simulation. Variation of parton distribution functions and uncertainty in the total integrated luminosity result in additional 4\% and 6.1\% errors in signal yield respectively. 
The total uncertainty on the background is dominated by statistics.

\begin{table}
\begin{ruledtabular}
\begin{tabular}{cccccc}
$\Lambda$, TeV &  $m_{\tilde{\chi}_1^0}$, GeV & $m_{\tilde{\chi}_1^+}$, GeV
     & $\sigma^{LO}$, fb & $k$-factor & Efficiency \\ \hline
 70  &  93.7 & 168.2 & 215  & 1.21 & $0.17\pm 0.03$ \\
 75  & 101.0 & 182.3 & 148    & 1.20 & $0.18\pm 0.03$ \\
 80  & 108.5 & 198.1 & 97.5 & 1.19 & $0.18\pm 0.03$ \\
 85  & 115.8 & 212.0 & 65.4 & 1.18 & $0.19\pm 0.03$ \\
 90  & 123.0 & 225.8 & 41.8 & 1.17 & $0.19\pm 0.03$ \\
 95  & 130.2 & 239.7 & 29.5 & 1.16 & $0.20\pm 0.03$ \\
 100 & 137.4 & 253.4 & 20.6& 1.15 & $0.20\pm 0.03$ \\
 105 & 144.5 & 267.0 & 14.4& 1.14 & $0.18\pm 0.03$ \\
 110 & 151.7 & 280.7 & 10.3& 1.13 & $0.19\pm 0.03$ \\
\end{tabular}
\caption{ \label{tab:susy}
Points on the GMSB Snowmass Slope model: neutralino and chargino
masses, cross sections predicted by {\sc PYTHIA}, $k$-factors, and reconstruction efficiencies with total uncertainty.}
\end{ruledtabular}
\end{table}


As the observed number of events for all values of \MET is in good agreement
with the standard model prediction, we conclude that there is no evidence for GMSB SUSY
in the data. We set limits on the production cross section by utilizing a likelihood
fitter~\cite{likelihoodFitter} that incorporates a log-likelihood ratio
(LLR) test statistic method. This method utilizes binned \MET distributions
rather than a single-bin (fully-integrated) value, and therefore accounts
for the shapes of the distributions, leading to greater sensitivity.
The value of the confidence level for the signal
$CL_s$ is defined as $CL_s = CL_{s+b}/CL_b$, where $CL_{s+b}$ and $CL_b$ are the
confidence levels for the signal plus background hypothesis and the
background-only (null) hypothesis, respectively. These confidence levels
are evaluated by integrating corresponding LLR distributions populated
by simulating outcomes via Poisson statistics. Systematic uncertainties are
treated as uncertainties on the expected numbers of signal and background
events, not the outcomes of the limit calculations. 
The degrading effects of systematic uncertainties are reduced by introducing 
a maximum likelihood fit to the missing transverse energy distribution.  A separate 
fit is performed for both the background-only and signal-plus-background hypotheses 
for each data or pseudo-data distribution.

\begin{figure}
  \includegraphics[width=9cm]{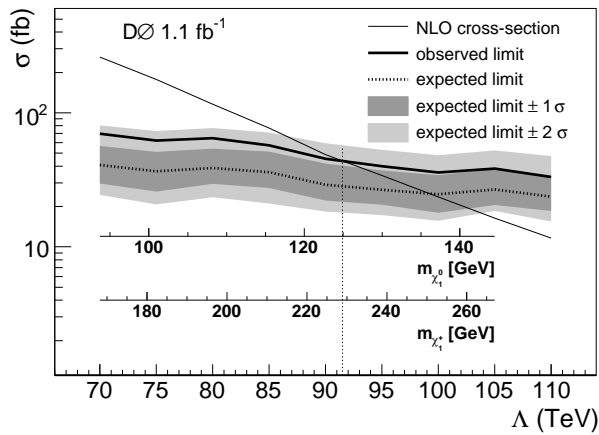}\\
  \caption{Predicted cross section for the Snowmass Slope model versus $\Lambda$. The observed and expected 95\% C.L. limits are shown in
  solid and dash-dotted lines, respectively.
  }\label{fig:limit}
\end{figure}

The limits are shown in Fig.~\ref{fig:limit} together with
expected signal cross sections. The observed limits are statistically compatible
with the expected limits. The observed upper limit on the signal cross section
is below the prediction of the Snowmass Slope model for $\Lambda < 91.5$ TeV, or in terms of
gaugino masses, $m_{\tilde{\chi}_1^0} < 125$~GeV and $m_{\tilde{\chi}_1^+} < 229$~GeV.
These represent the most stringent limits on this particular GMSB SUSY model to date.

%
We thank the staffs at Fermilab and collaborating institutions, 
and acknowledge support from the 
DOE and NSF (USA);
CEA and CNRS/IN2P3 (France);
FASI, Rosatom and RFBR (Russia);
CAPES, CNPq, FAPERJ, FAPESP and FUNDUNESP (Brazil);
DAE and DST (India);
Colciencias (Colombia);
CONACyT (Mexico);
KRF and KOSEF (Korea);
CONICET and UBACyT (Argentina);
FOM (The Netherlands);
Science and Technology Facilities Council (United Kingdom);
MSMT and GACR (Czech Republic);
CRC Program, CFI, NSERC and WestGrid Project (Canada);
BMBF and DFG (Germany);
SFI (Ireland);
The Swedish Research Council (Sweden);
CAS and CNSF (China);
Alexander von Humboldt Foundation;
and the Marie Curie Program.
%

\end{document}